\documentclass[review]{elsarticle}

\usepackage{lineno,hyperref}
\modulolinenumbers[5]
\usepackage{amsmath}

\journal{Journal of Nuclear Instruments and Methods in Physics Research A}

\begin{document}

\begin{frontmatter}

\title{A 100-ps Multi-Time over Threshold Data Acquisition System for Cosmic Ray Detection\tnoteref{mytitlenote}}
\tnotetext[mytitlenote]{The first two authors contributed equally to the paper and are listed in alphabetical order.}


\author[mymainaddress]{Konstantina Georgakopoulou}

\author[mymainaddress]{Christos Spathis}

\author[mysecondaryaddress]{Georgios Bourlis}

\author[mysecondaryaddress]{Apostolos Tsirigotis}

\author[mymainaddress]{Alexios Birbas\corref{mycorrespondingauthor}}
\cortext[mycorrespondingauthor]{Corresponding author}
\ead{birbas@ece.upatras.gr}

\author[mysecondaryaddress]{Antonios Leisos}

\author[mymainaddress]{Michael Birbas}

\author[mysecondaryaddress]{Spyros E. Tzamarias}

\address[mymainaddress]{Department of Electrical and Computer Engineering, University of Patras, Patras, 26500 Greece}
\address[mysecondaryaddress]{Physics Laboratory, School of Science \& Technology, Hellenic Open University, Tsamadou 13-15 Patras, 26222, Greece}

\begin{abstract}
High-energy cosmic rays are one of the primary sources of information for scientists investigating the elementary properties of matter. The need to study cosmic rays, with energies thousands of times larger than those encountered in particle accelerators, led to the development of modern detection hardware and experimental methodologies. We present a low power, low complexity data acquisition (DAQ) system with 100 ps resolution, suitable for particle and radiation detection experiments. The system uses a Multiple-Time-over-Threshold (MToT) technique for the treatment of the output signal of Photo Multiplier Tubes (PMTs). The use of three thresholds compensates for the slewing effects and offers a more accurate measurement of the PMT pulses' width. For the evaluation of the pulse the system uses comparators and a Time-to-Digital (TDC) converter, whereas the pulses are time-stamped using the GPS signal. The prototype card is analyzed for its noise behavior and is tested to verify its performance. The system has been designed for the HEllenic LYceum Cosmic Observatories Network (HELYCON) Extensive Air Showers (EAS) detector.
\end{abstract}

\begin{keyword}
Multiple-Time-over-Threshold, Data Acquisition card, HELYCON
\end{keyword}

\end{frontmatter}


\section{Introduction}
Until the middle of the 20th century, high-energy cosmic rays were the primary source of information for scientists investigating the elementary properties of matter. They were later superseded by particle accelerators that offered the means of conducting experiments with beams of energetic particles, paving the way for great progress in understanding the structure and interactions of matter. In recent years, there has been a resurgence of cosmic ray experiments; indeed, the development of modern detection hardware and experimental methodologies has made the study of cosmic rays, with energies thousands of times larger than those encountered in particle accelerators, a reality. 

These ambitious research programs would not have been possible without a constant development in the field of detection units and data acquisition (DAQ) hardware. The hardware units designed for such applications have to operate reliably and continuously, for extended periods of time and under the harshest environments, including inside ice, deep-sea underwater, etc. At the same time, they have to be advanced enough to offer large sensitivities, low-noise operation and be able to distinguish the useful signal from background noise, which is often many orders of magnitude higher in rate.

In this work, we describe a DAQ system suitable for particle and radiation detection experiments, involving Photo Multiplier Tubes (PMTs) as sensing devices. The prototype card was designed as part of the ASTRONEU project \cite{astroneu}, to function as a detection unit in a cosmic ray telescope aiming to detect Extensive Atmospheric Showers (EAS). The particle sensing units are comprised of plastic scintillators connected to PMTs. The scintillator produces photons in response to incident ionizing particles. The photons are forwarded to the PMTs through optical fibers, where they get transduced to an electrical pulse; this pulse is the input of the DAQ system.

The developed prototype system can find use even outside the field of cosmic ray detection experiments. Even though the main focus of such astroparticle experiments is on cosmic rays with energies in the TeV to PeV range, such as high energy charged particles and neutrinos of astrophysical origin, the low energy threshold and the increased resolution of the DAQ unit can also prove useful in the detection of nuclear reactor neutrinos, as well as geo-neutrinos, among others.

Simulation studies have shown that a single threshold level is not enough to accurately measure the PMT pulses' characteristics. Furthermore, the effect of the pulse's waveform slewing is more pronounced in a single threshold system. The positive effects of using multiple thresholds have been detailed in \cite{MToT}, in a study investigating Time-over-Threshold (ToT) readout systems with up to 6 thresholds. The results show that, the superposition of multiple pulses, arising from delayed photons, leads to irregular waveform shapes at the PMTs output. The deviation of the PMTs' waveform from the standard shape corresponding to a single photoelectron can best be accounted for by using more than a single threshold; in fact, at least three threshold levels would be necessary to enable a detailed pulse reconstruction. The use of three thresholds is also adequate to compensate for the slewing effects and aids in the accurate reconstruction of the event's energy by estimating the waveform's charge and thus the number of photons incident on the PMT. The reconstruction of the event's energy is accomplished comparing this number of photons with the expected one estimated with Monte Carlo simulation.

The data acquisition systems typically used for cosmic ray detection applications undertake the tasks of interfacing with the PMTs, digitizing the input signal, time-stamping the recorded events and sending the processed data to a computer. The digitization of the input signal can be achieved in various ways, most often by using a traditional ADC architecture, like a flash ADC.


\section{Time over Threshold}
\begin{figure}[!t]
	\centering
	\includegraphics[width=3in]{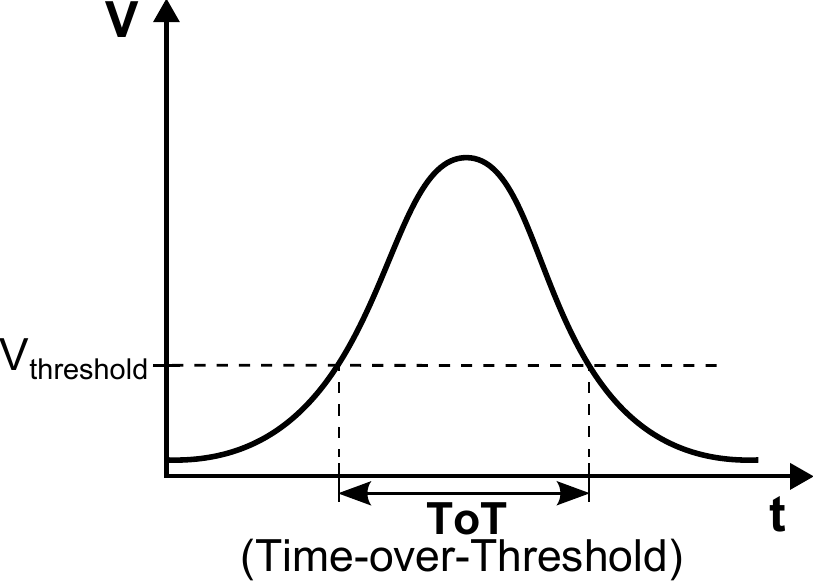}
	\caption{\label{fig:Figure_1} In the ``Time-over-Threshold'' digitization technique, the duration for which the input signal has values above a voltage threshold level is measured.}
\end{figure}

\begin{figure}[!t]
	\centering
	\includegraphics[width=3in]{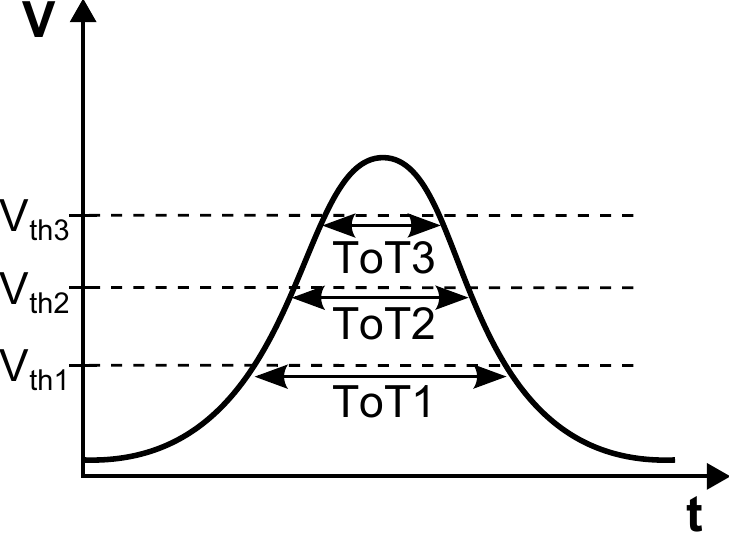}
 	\caption{\label{fig:Figure_2} The resolution of the ``Time-over-Threshold'' technique can be enhanced, by increasing the number of thresholds (Multiple ToT). The case of three thresholds is depicted, as used in the DAQ system of this work.}
\end{figure}

An interesting alternative method of digitization has been proposed in the literature, under the name of ``Time-over-Threshold'' \cite{ToT,ToTtzamaria}. The ToT method takes advantage of the fact that the shape of the PMT output pulses is, to some extent, predictable: the pulses can be approximated as standard shape curves or a superposition of many such curves. Thus, a time-based technique of digitizing the input signal can be realized by shifting the focus on how long a given pulse spends above a (user-predetermined) voltage threshold and extrapolating the rest of the shape (Fig.~\ref{fig:Figure_1}). This duration is defined by the output of a comparator (in this context also referred to as a``discriminator'') which compares the input signal to the threshold voltage. Then, the duration of the comparator's output pulse is digitized using a Time-to-Digital Converter (TDC). The individual pulses can be time-stamped using a GPS module, synchronized to the TDC output data. Finally, the resulting ToT duration can be used to extract information about the approximate shape and therefore, charge (area under the curve) of each event. It is easy to see that, increasing the number of thresholds used, leads to a direct enhancement of the digitization resolution (Fig.~\ref{fig:Figure_2}). The ToT method with multiple thresholds is known as ``Multiple ToT'' or ``MToT''.

The advantages of using a time-based digitization method such as the ToT over traditional ADC architectures are many, but the most significant are the low power consumption and the reduction of complex accessory subcircuits. This combination results in an increase of the reliability of the overall system, an attribute very important for detection units that are to be constantly online or are difficult and costly to repair, e.g. modules that operate in ice or underwater, at great depths. In such harsh environments (high pressure, salt water, low temperatures, etc.), it is vital for the electronics designer to minimize the chances of failure. In addition, the ToT technique can offer, in such applications, comparable performance to typical, fast ADC architectures (flash, SAR, pipeline) at a reduced price. 

\subsection{Architecture Overview}
\begin{figure}[!t]
	\centering
	\includegraphics[width=4.5in]{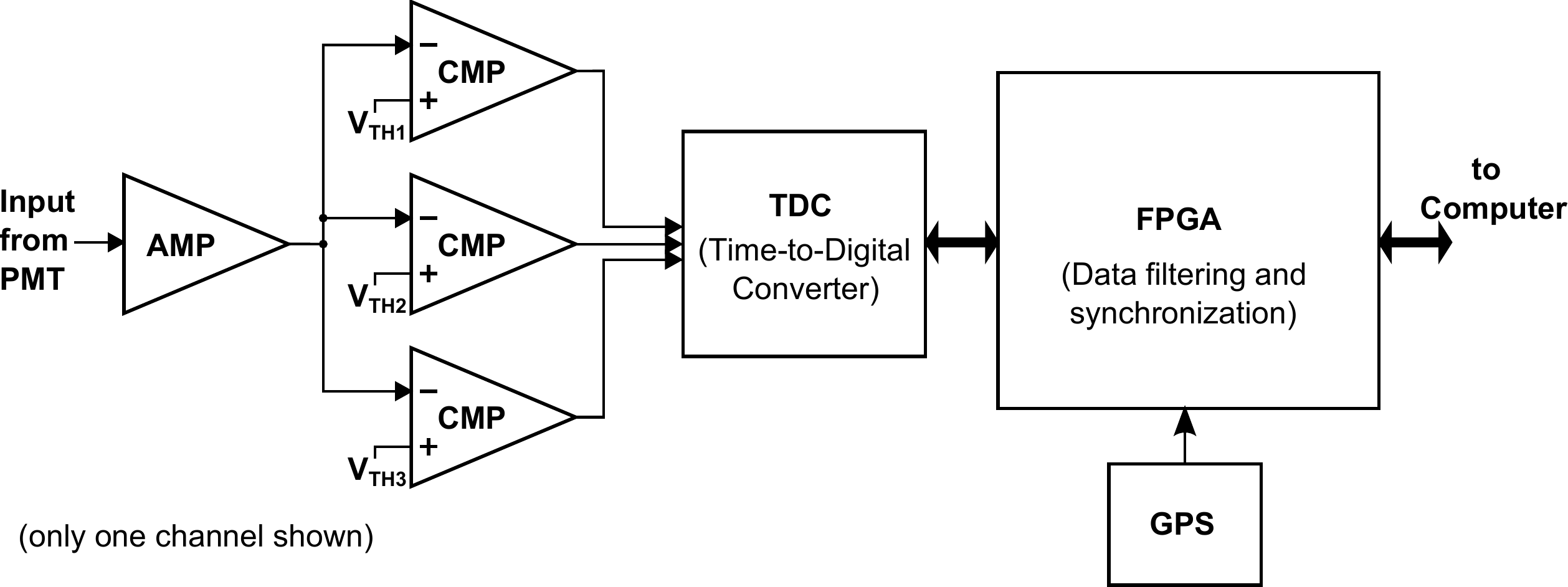}
	\caption{\label{fig:Figure_3} Overview of the Cosmic Ray detection system architecture. The input signal from the PMT is first buffered or amplified, then digitized via the three threshold MToT method. To that end, three comparators are used, one for each threshold. The ToT measurement is handled by the TDC. An FPGA processes the acquired measurement data, whereas a GPS module serves the function of synchronizing events from different detection stations, geographically separated. Only one channel is shown.}
\end{figure}

The general architecture of the developed DAQ system is shown in Fig. ~\ref{fig:Figure_3}. The system was designed to be compatible with any cosmic ray experiment setup that utilizes PMTs, as part of its sensing device; nevertheless, due to the prototype card's size, it would be better suited to experiments where the front-end electronics need not be integrated inside an Optical Module, also housing the PMTs, and therefore, low area is not a high priority (e.g. as in the HELYCON experiment \cite{helycon, HELYCON2}). As mentioned before, the system makes use of the MToT readout technique.

The operation of the DAQ can be outlined as follows: the particle detector, including the PMT converts an incoming optical event (e.g. the passage of a muon through a scintillator) to an electrical pulse. This pulse reaches the input of the DAQ card, where it is buffered or amplified, depending on the user's preference. Three comparators are used to implement an MToT technique with three thresholds, all of which are user-adjustable. The comparator outputs are connected to a special purpose TDC, which is tasked with measuring and digitizing the duration of the pulses the comparators produce at their output. Next, the data from the TDC are passed on to an FPGA, which handles their filtering and initial processing. Finally, the processed, time-stamped data are sent to a computer. The GPS module connected to the prototype card offers data synchronization capabilities in both time and space: the events can be time-stamped by extracting the Coordinated Universal Time (UTC) information, whereas the data from various detection units are related to each other by providing their unique position coordinates. The information of the pulse arrival time and detector position can help determine the origin of the incoming showers, by calculating their angle of incidence. Moreover, the output ToT values can be used to estimate the number of particles that crossed the scintillator and help in the estimation of the energy of the cosmic particle that initiated the shower. The overall system is comprised of the following parts: the prototype DAQ card, the FPGA development board, the Altera DK-CYCII-2C20N and the GPS module, Leadtek LR9450.

\subsubsection{Analog front-end}     
The analog front-end schematic is comprised of an input amplifier, a comparator and a reference voltage circuit, for the generation of the threshold voltages. The amplifier at the card's input can be configured as either a buffer or a gain stage, with the gain selected by the user. The crucial role of the amplifier at the system's input places strict requirements on its noise performance. Perhaps the most important attribute, however, that the amplifier should possess is high bandwidth and slew rate. The typical input pulse has rise/fall times down to a few ns and its width varies between a few hundred ns down to less than 10 ns.

Based on the aforementioned requirements, a current feedback amplifier (CFB) was selected as the amplifying device. This class of amplifiers is often associated with high output current capabilities and high bandwidths that vary weakly with different gain configurations. The specific amplifier is an excellent choice for a pulse amplifier, having a bandwidth of 1 GHz, when operating as a buffer and slew rates approximately $SR$ = 5500 V/$\mu$s, for a capacitive load up to 50 pF: it can swing from 0 V to -5 V in less than 1 ns. The amplifier's offset is typically 2 mV, which is somewhat high, but can be calibrated out by adjusting the thresholds appropriately. The three thresholds are set individually by highly accurate and linear trimming potentiometers. The thresholds are driven by a temperature compensated, low-drift, precision voltage reference. 

The comparator needs to exhibit low rise/fall times and propagation delay, even with relatively low overdrives. Achieving this performance with anything other than an advanced-process ASIC is difficult. Even though a potential propagation delay will ultimately affect the time-stamping of the event (the ``absolute'' time), as long as this delay can be held at similar levels between different prototype cards, the ``relative'' time between events at different detection units will remain constant. Low propagation delays can be achieved by using enough gain in the amplifying stage to guarantee a minimum overdrive for the majority of the incoming events. Also, as long as the propagation delay is symmetric for rising and falling edges, the ToT measurement will be largely unaffected. The symmetry of the rise/fall times of the output, on the other hand, are more detrimental for an accurate ToT acquisition.

The typical rise and fall times of the comparator used in the prototype card are both 2 ns for a load of 10 pF. Also, the propagation delay is typically 4 ns with an overdrive of just 20 mV and a load of 10 pF. Given that the overdrive, at least for the first threshold, is much higher (or can be adjusted to be, through the gain setting) and that the comparator's load is closer to 6 pF, the expected performance should be even better.  

The description so far has focused on one channel of the DAQ system. The prototype card has two such channels, with three thresholds. In this version of the system, the thresholds are common for both channels, but an independently adjustable version is also planned for the future.

One could wonder whether or not the use of an integrated ADC, following more traditional digitization architectures, would be beneficial. To achieve the challenging specifications set by the cosmic ray detection purposes, one would have to select ``fast'' ADCs, such as a flash or a SAR ADC, etc. Unfortunately, such advanced architectures are much more expensive and consume much more power. The latter characteristic is alarming in two ways: first, low-power implementations become impossible to achieve and second, the increased heat dissipation can affect negatively the reliability of the overall system, especially if it is enclosed (as in an optical module) and heatsinks and/or fans are not available; high-end oscilloscopes that incorporate such ADC ICs have powerful fans to dissipate the generated heat. 

More specifically, the power drawn from the selected TDC is 160 mW (maximum) and its price is at 138\$. At the time of writing, commercial ADCs with sampling rates in the 500 MS/s to 2.5 GS/s range, dissipate at least 1.38 W (4.6 W on average) and cost more than 370\$ (880\$ on average), for two channel implementations as in the prototype card. If the necessity for accessory components, specialized differential driver-amplifiers, etc. is factored in, then the benefits of using a time-based digitization technique, such as the ToT, becomes even more evident. Consequently, even though an integrated ADC can, in theory, offer superior performance, in practice and for this specific application, the price paid is quite high and the difference in the digitization fidelity not important enough to discourage the use of the MToT method.

\subsubsection{Digital subcircuit}
The digital part of the system consists of the TDC and the Altera Cyclone II FPGA, along with its development board, for this version of the DAQ system. The TDC is responsible for the accurate measurement of the duration of the comparator's output pulses (ToTs). Furthermore, it handles the task of time-stamping every event recorded, by measuring the time difference between the arrival of an event and the 1PPS signal supplied by the GPS. Thus, the TDC provides both the duration of each ToT and its place in UTC timeline. The measurement resolution of the ToT is typically 100 ps.
 
The FPGA is the main data processing unit of the overall system. The TDC is programmed and accessed through the FPGA. In addition, any post processing steps and filtering tasks are performed here: the data are filtered for noise and the ``coincidence algorithm'' can be used, if necessary. The former entails the rejection of pulses that are too long or too short (e.g. greater than 1 $\mu$s) to be valid events. The latter checks for the offset in the time of arrival between events captured by the two channels, since each event is time-stamped via the GPS. If this offset is less than a user specified length (e.g. 500 ns), the event can be considered valid with some confidence, otherwise it is considered noise and is discarded. Finally, it offers RS232 communication between the system and the computer.

\begin{figure}[!t]
	\centering
	\includegraphics[width=3in]{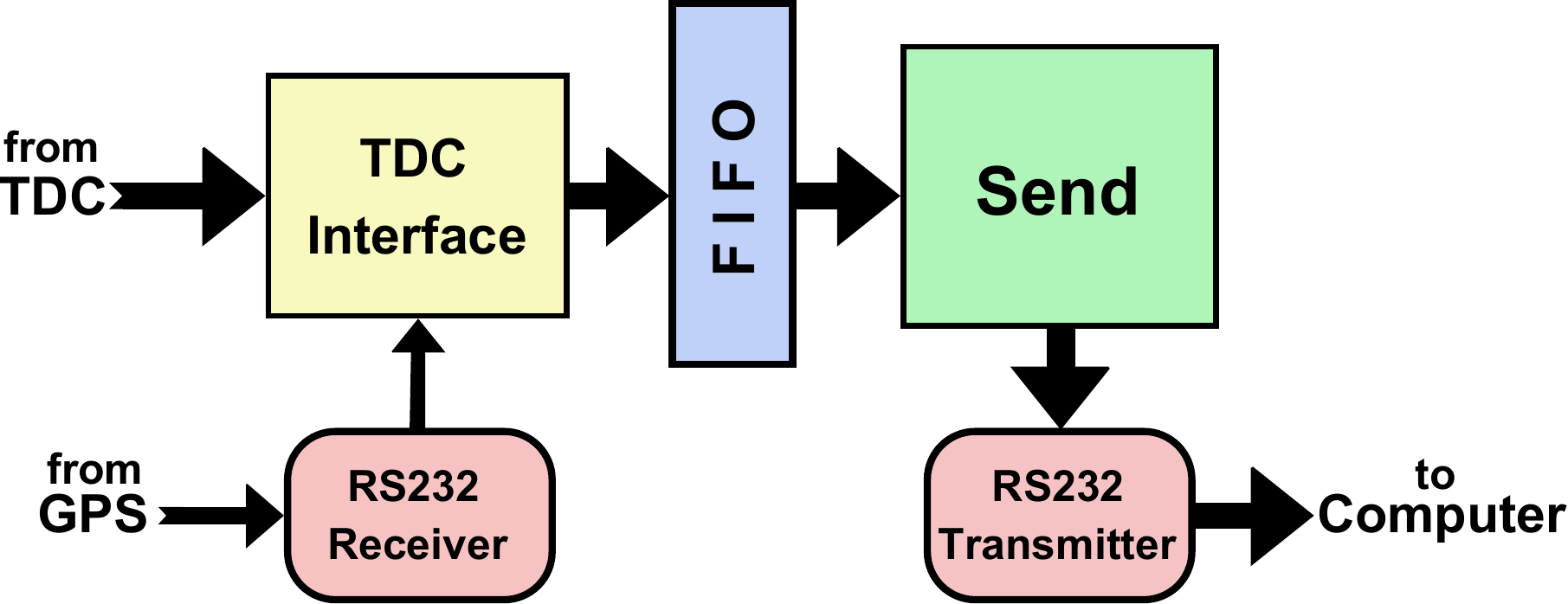}
	\caption{\label{fig:Digital} The schematic of the digital processing part of the system.}
\end{figure}

The schematic of the digital processing part of the system is shown in Fig. ~\ref{fig:Digital}. The digital section, implemented in the FPGA, contains blocks of different functionalities: the interfacing with the TDC (programming and data reading), a FIFO, the data processing (filtering and data transformation into ASCII format) and RS232 receiver/transmitter modules for communicating with the GPS and the computer. 

The ``TDC Interface'' block handles the interfacing tasks associated with the TDC. It is responsible for initially programming the converter. Under normal operation, it continually reads the measurement data of the TDC at a fast rate (up to 140 MHz) and passes it on to a FIFO register. 

The FIFO is used to temporarily store the data received; this allows the TDC Interface to operate at relatively high frequencies, even though the data is processed and filtered at a slower rate, before they are sent to the computer. The FIFO register is large enough (1024 x 195) to handle the maximum continuous event triggers expected, i.e. during a particle ``shower''. 

An RS232 receiver block reads the Recommended Minimum specific GPS data (RMC message), which contains an ASCII message with information such as the date, time, position, satellite connection etc. This message is updated every second and appended to the data received during that time in the TDC Interface block, before storing it in the FIFO.

The ``Send'' block converts the data into ASCII format. Prior to this, any computation necessary to acquire the ToT offset and duration is performed here. In addition, it filters the measurements for noise and coincidence level, as discussed earlier. Finally, the completed message is passed on to an RS232 transmitter, to be sent to the computer.

Note that the schematic only aims in describing the main flow of information in the digital domain. Any handshaking signals, such as the FIFO empty/full flags flowing in the reverse direction (i.e. from the FIFO to the TDC Interface block), are omitted for clarity.

The current version of the DAQ system uses a commercial TDC to measure the ToT durations. This choice was made on the grounds of achieving better reproducibility and stability of the measurements under different operating conditions, since the internal voltage references and oscillators of the TDC chip are tightly regulated. Nevertheless, an additional single-threshold channel is connected to the FPGA directly, in order to perform research on the possible implementation of the TDC functionality inside the FPGA, as has been suggested and described elsewhere \cite{TDCinFPGA}. This approach, although versatile and very efficient, typically suffers from inconsistencies in the measurement results, compared to an ASIC. However, given the large amount of channels that can be connected to a single FPGA, as well as the convenience of using a single FPGA for both measuring and processing the ToT durations, it is an alternative worth exploring. The research in the literature towards that direction is constantly expanding \cite{TDC1,TDC2,TDC3}.

\subsubsection{The prototype card}
\begin{figure}[!t]
	\centering
	\includegraphics[width=4in]{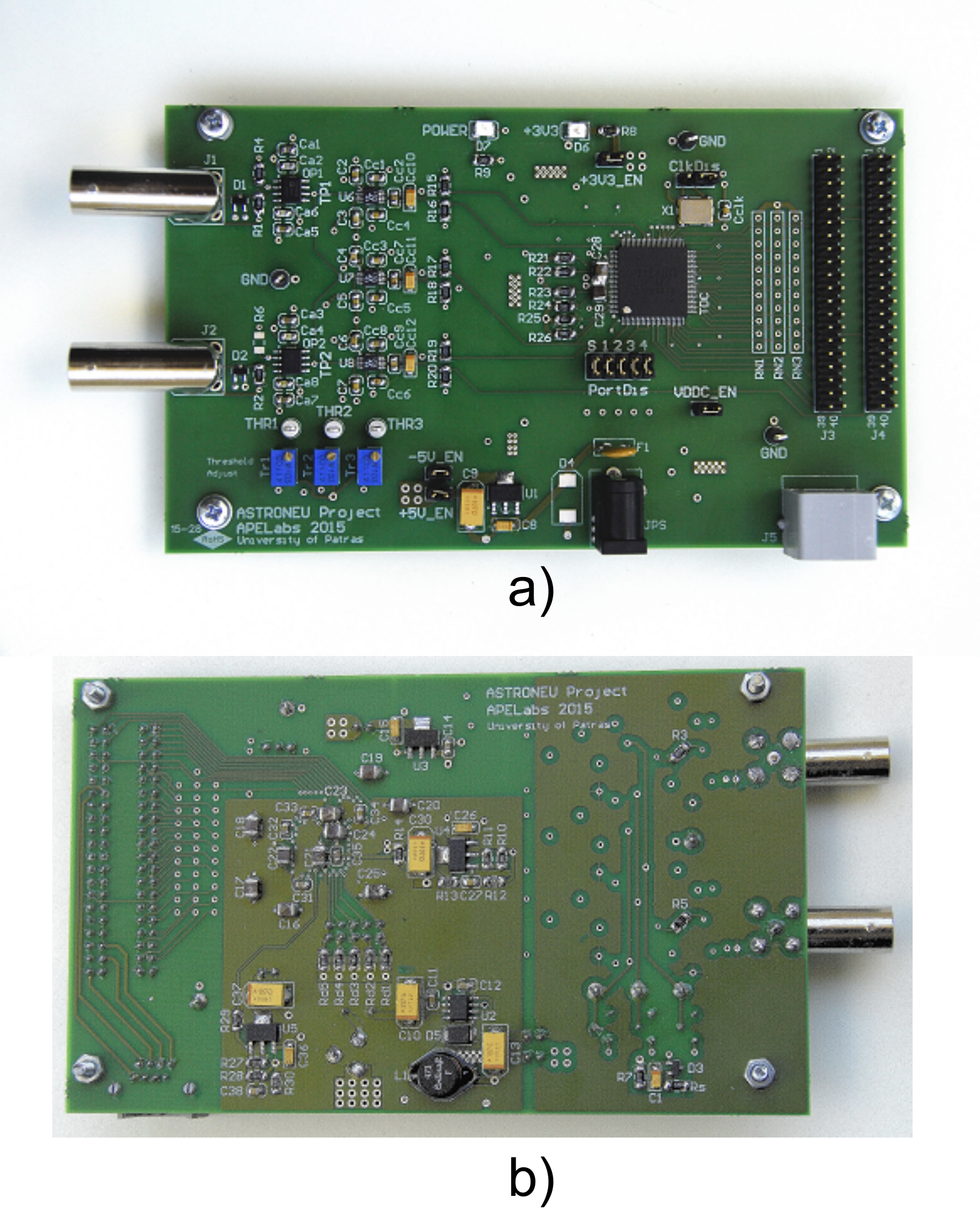}
	\caption{\label{fig:EAP_PCB} The prototype card: the card's top (a) and bottom side (b).}
\end{figure}

The prototype PCB designed contains four layers: two signal layers (top and bottom) and two power planes (in between). Figure ~\ref{fig:EAP_PCB} a and b demonstrate the top and bottom side of the PCB, respectively.

The various specifications and figures-of-merit of the prototype card are summarized in Table~\ref{tbl:EAP}.

\begin{table}
\begin{tabular}{ |p{5cm}||p{5cm}| }
 \hline
 Channel number & $2$\\
 Resolution & $0.1 ns$\\
 Minimum pulse duration & $<1 ns$\\
 Threshold number & $3$, individually adjustable\\
 Input voltage range & $-5 V$ to $+5 V$\\
 Power Supply & $7.5V$ DC\\
 Power consumption & $1.5 W$ (max)\\
 PCB Layers & $4$ ($2$ signal + $2$ power)\\
 Card dimensions & $146 mm$ $\times$ $87 mm$\\
 \hline
\end{tabular}
\caption{Summary of the prototype card specifications}
\label{tbl:EAP}
\end{table}

\subsection{Noise analysis}
As in most time-based circuits that make use of comparators, the voltage noise present at the input of the comparator is ultimately transduced to noise in the timing phase (jitter) of the comparator output pulse, i.e. the ToTs measured. To quantify the effect of this phenomenon for this work, the findings of \cite{NoiseComparator} are used. Before proceeding, it is worth mentioning that, due to the nature of the detection experiment and the high frequencies involved, the major noise contribution is from ``white'' noise sources; low-frequency sources will be ignored hereafter.

\begin{figure}[!t]
	\centering
	\includegraphics[width=4in]{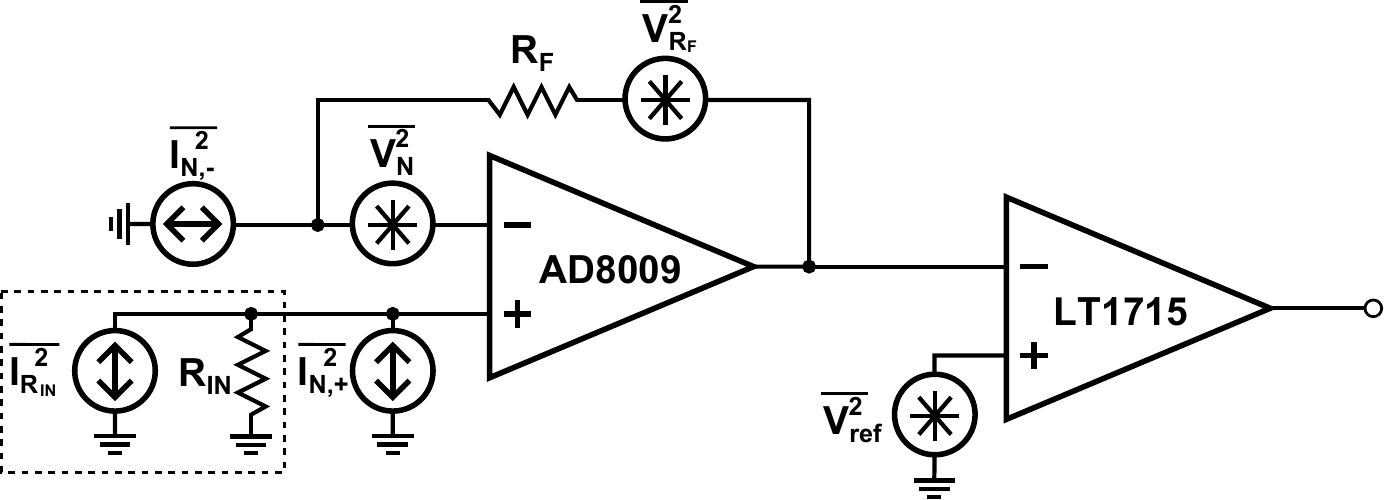}
	\caption{\label{fig:EAPopampNoise} Noise analysis circuit of the DAQ system. The amplifier's noise is given by the sources $\overline{V_N^2}$, $\overline{I_{N,+}^2}$ and $\overline{I_{N,-}^2}$. The noise due to the resistors is represented by $\overline{I_{R_{IN}}^2}$ and $\overline{V_{R_F}^2}$. The voltage noise $\overline{V_{ref}^2}$ is due to the reference voltage generator. In this analysis the amplifier is used as a buffer and the resistor $R_G$ can be omitted.}
\end{figure}

The voltage noise at the comparator's input originates in the amplifier's output, the comparator's internal preamplifying stage and the threshold generation circuit. Starting from the amplifier, the equivalent noise circuit is shown in Fig. ~\ref{fig:EAPopampNoise}. The noise sources are: the voltage noise of the feedback resistor $\overline{V_{R_F}^2}$, the current noise of the input resistor $\overline{I_{R_{IN}}^2}$, the input-referred voltage noise of the amplifier $\overline{V_N^2}$ and the input-referred current noise of the amplifier's input terminals, $\overline{I_{N,+}^2}$ and $\overline{I_{N,-}^2}$, for the non-inverting and inverting terminal, respectively.

The overall voltage noise is computed at the output, $V_{o,RMS}$. The noise gain (NG) is the gain of the amplifier in the non-inverting configuration and for this analysis, we assume $NG$ = 1. Also, the equivalent noise bandwidth is $EQBW$ = 1.57 BW, in this case. From the data sheet of the amplifier and for the resistor values shown, the noise densities are:
\begin{align}
\sqrt{I_{N,+}^2(f)} 		&= 46  pA/\sqrt{Hz}  \\
\sqrt{I_{N,-}^2(f)} 		&= 41  pA/\sqrt{Hz}  \\
\sqrt{V_N^2(f)} 			&= 1.9 nV/\sqrt{Hz}  \\
\sqrt{V_{R_F}^2(f)} 		&= 2.2 nV/\sqrt{Hz}  \\
\sqrt{I_{R_{IN}}^2(f)} 		&= 18  pA/\sqrt{Hz},
\end{align}
where the first three values are evaluated at $f$ = 10 MHz. Measuring the noise at this frequency guarantees that the results are not affected by the presence of low-frequency noise.  

The voltage noise at the amplifier's output can now be given from:
\begin{equation}
\label{eqn:EAPOutputNoise}
V_{o,RMS}^2=\left[ \overline{V_N^2}+\overline{V_{R_F}^2}+\overline{I_{N,-}^2}R_F^2+\left( \overline{I_{R_{IN}}^2}+\overline{I_{N,+}^2} \right)R_{IN}^2 \right]EQBW .
\end{equation}
The current noise of the inverting input terminal $I_{N,-}$, which becomes a voltage noise across $R_F$, contributes the most to the overall noise. The resulting noise is:
\begin{equation}
\label{eqn:EAPOutputNoise2}
V_{o,RMS} \approx 512\mu V .
\end{equation}
This noise value is relatively high, due to the very wide bandwidth and the gain configuration chosen. Including a gain resistor $R_G$ (for which typically $R_G < R_F$) would reduce the portion of $\overline{I_{N,-}}$ appearing at the output, lowering the overall noise.

The reference voltage noise is given as $V_{ref,RMS} \approx$ 35 $\mu$V, in the given bandwidth. Summing the noise calculated thus far at the comparator's input in an RMS fashion, we get:
\begin{equation}
\label{eqn:EAPOutputNoise3}
V_{in,RMS} \approx 513.2\mu V ,
\end{equation}
which clearly shows that the amplifier noise dominates.

The slope at the input of the comparator chosen \cite{LT1715} can be approximated as:
\begin{equation}
\label{eqn:EAPOutputNoise4}
\frac{dV}{dt}\vert _{min} \approx 48V/\mu s .
\end{equation}
Therefore, the jitter at the comparator's output, due to the card's input stage alone (amplifier and voltage reference), is:
\begin{equation}
\label{eqn:EAPOutputNoise5}
\sigma _{t_{in},RMS} = V_{in,RMS}\frac{1}{dV/dt} \approx 10.7 ps .
\end{equation}

The jitter of the comparator itself (due to noise from its internal preamplifier) is given separately for positive and negative transitions of its output, as $\sigma _{+,RMS}$ = 15 ps and $\sigma _{-,RMS}$ = 11 ps, respectively. Since the system digitizes pulses that cause both a positive and a negative transition of the output, we can RMS-sum the jitter contributions of the comparator's input stage as a worst case scenario, as if they were uncorrelated noise sources - they are not, since they both originate in the same preamplifying circuit. Then, the overall output jitter becomes:
\begin{equation}
\label{eqn:EAPOutputNoise6}
\sigma _{t,RMS} \approx 21.5 ps .
\end{equation}
This noise is small and cannot meaningfully affect the measurement in any way. To prove this, consider that the TDC resolution alone is higher than the jitter, at $\approx$ 81 ps. What is more, this specific application has no need for a resolution much better than 1 ns, due to the nature of the pulses processed.

\section{Results and Discussion}
For the calibration and evaluation of the card, a Tektronix AFG3252C arbitrary function generator has been employed, to produce pulses with characteristics (shape, rise and fall time) similar to the ones from HELYCON scintillator detectors. The signal is split and led to the two inputs of the card and to two channels of a high sampling rate oscilloscope, namely a Tektronix DPO 7104C one. The card prototype is triggered independently for each channel, based on the lower threshold value set for each one. 

\begin{figure}[!ht]
	\centering
	\includegraphics[width=3in]{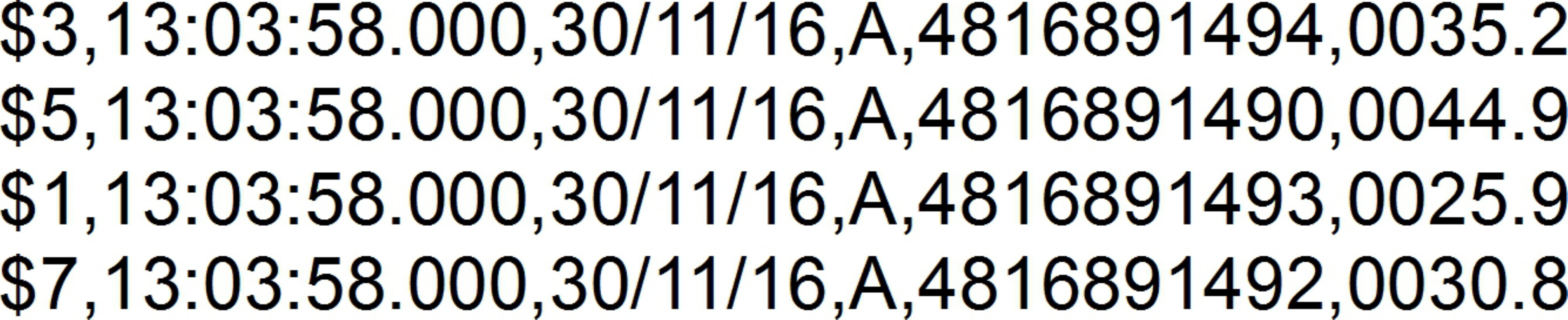}
	\caption{\label{fig:Fig8} Format of the data transmitted by the card.}
\end{figure}

Regarding the communication with the card and data recording, Hyperterminal has been employed. The format of the data transmitted by the card is depicted in Fig. ~\ref{fig:Fig8}. The first field encodes the channel and threshold number, with \$3 indicating data from the first (and only) threshold of the second channel, while \$5, \$7 and \$1 mark data from the first, second and third threshold respectively of the first channel. The next two fields are the time and date of the event, as supplied by the GPS module, while the field following them indicates whether the GPS data are valid. The following field is the time duration (in nanoseconds) of each threshold crossing of the event with respect to the GPS time. The time value corresponding to the first (lower) threshold of the channels can be used for the timing of the pulses. This, of course, introduces an error in the timing of the pulses that depends on their size (slewing), but this is not dealt with in this work, since identical pulses have been used for the calibration purposes. The last field is the ToT values, i.e. the time duration that the pulses of each channel remained above the respective threshold.

On the oscilloscope side full digitization of the waveforms is performed, with a sampling rate of 5 GS/s, thus recording waveforms with a time bin of 0.2 ns. A large number of data is collected by both the oscilloscope and the card, and analyzed in order to obtain an exact relation between the set thresholds (using the card potentiometers, Bourns 3266W) and the real operating values. Any offsets in the threshold values are mainly due to the amplifier and the comparators. These offsets are static and can be easily calibrated out from the measurements. This is achieved by calculating the ToT values from the oscilloscope data at various thresholds and comparing them with the ToT values recorded by the card.

\begin{figure}[!ht]
	\centering
	\includegraphics[width=3in]{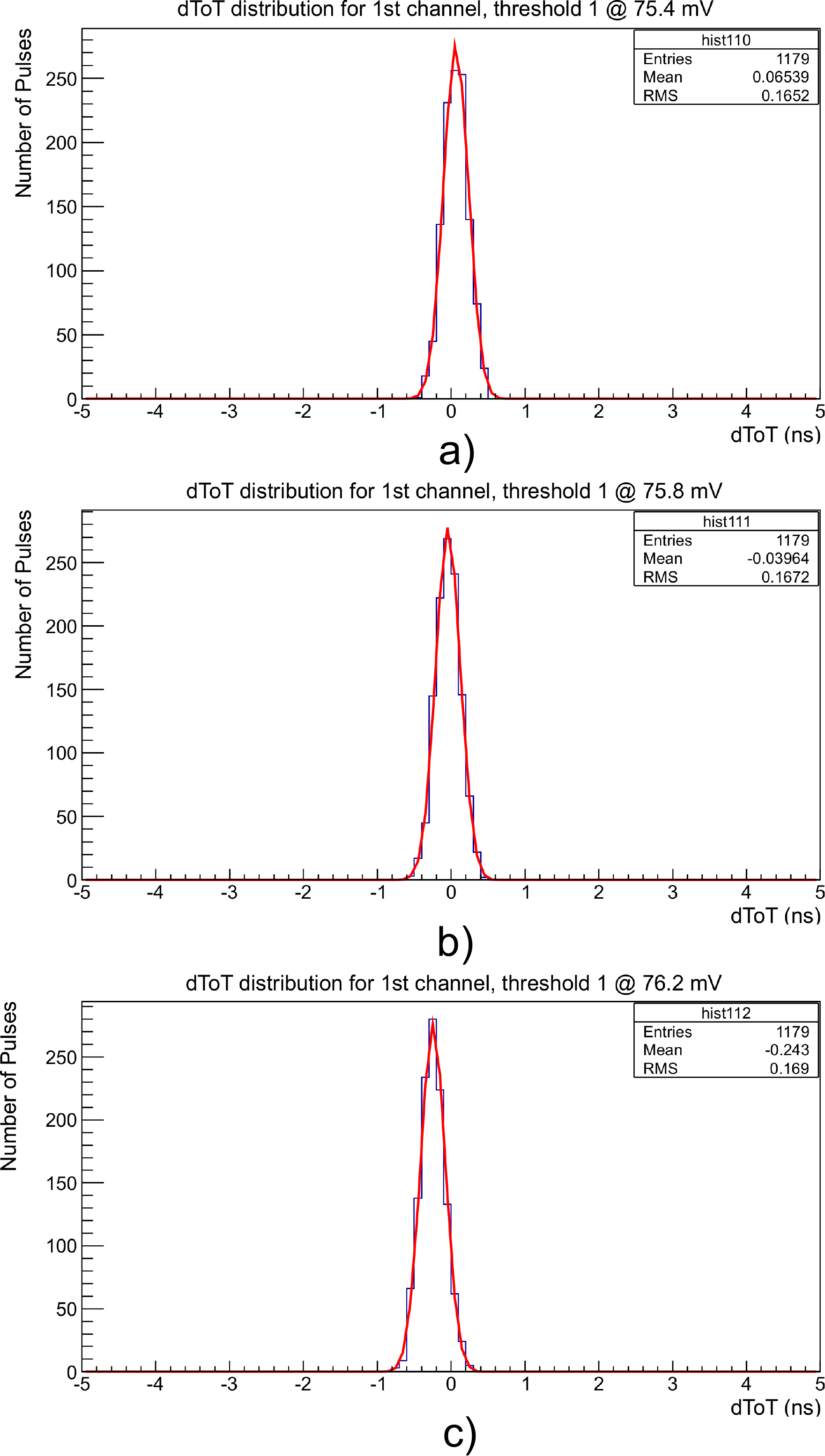}
	\caption{\label{fig:Fig9} Distribution of the difference of the card ToT value and the oscilloscope mean ToT value as calculated by the fully digitized waveforms. Each distribution corresponds to a different amplitude level employed to calculate the ToT values from the oscilloscope data. These are 75.4 mV, 75.8 mV and 76.2 mV from top to bottom respectively. The card threshold in this example was set to 80 mV.}
\end{figure}

From the fully digitized waveforms of the oscilloscope, the Time over Threshold (ToT) values are calculated at a number of thresholds values around the set value, with a certain step, typically 0.4 mV. For each threshold the average and the sigma of the ToT values are calculated. Subsequently, from the card data, the ToT value is obtained and the distribution of the difference of this value and the respective mean value from the oscilloscope data (dToT in the following) is generated for each threshold value. The mean value of the distribution of dToT should be zero for the threshold that corresponds to the actual threshold that has been applied to the respective channel of the card. Such dToT distributions are depicted in Fig. ~\ref{fig:Fig9} for the first channel of the card for a set threshold value of 80 mV. For the three distributions depicted the employed threshold for the ToT calculation from the oscilloscope data is 75.4 mV, 75.8 mV and 76.2 mV, respectively. The threshold value that equals the actual value of the threshold set on the channel should give a distribution with mean zero, which is almost the case for the second distribution in Fig. ~\ref{fig:Fig9}.

\begin{figure}[!t]
	\centering
	\includegraphics[width=4in]{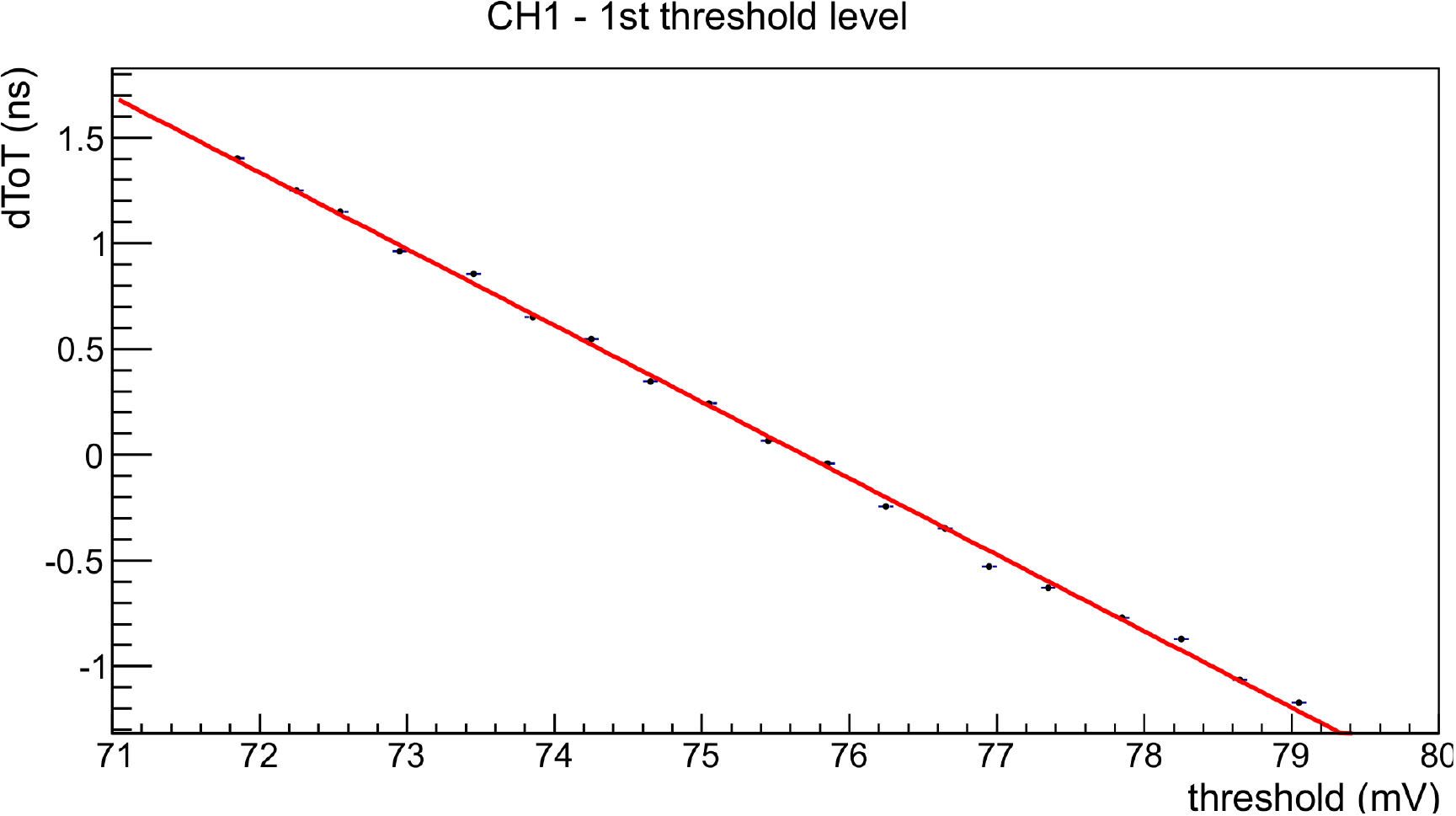}
	\caption{\label{fig:Fig10} Mean value of the distribution of dToT versus the employed threshold on the oscilloscope for the 1st threshold of channel 1. The threshold value of the line intersection with the x axis, is the actual threshold value set on the card.}
\end{figure}

\begin{figure}[!t]
	\centering
	\includegraphics[width=4in]{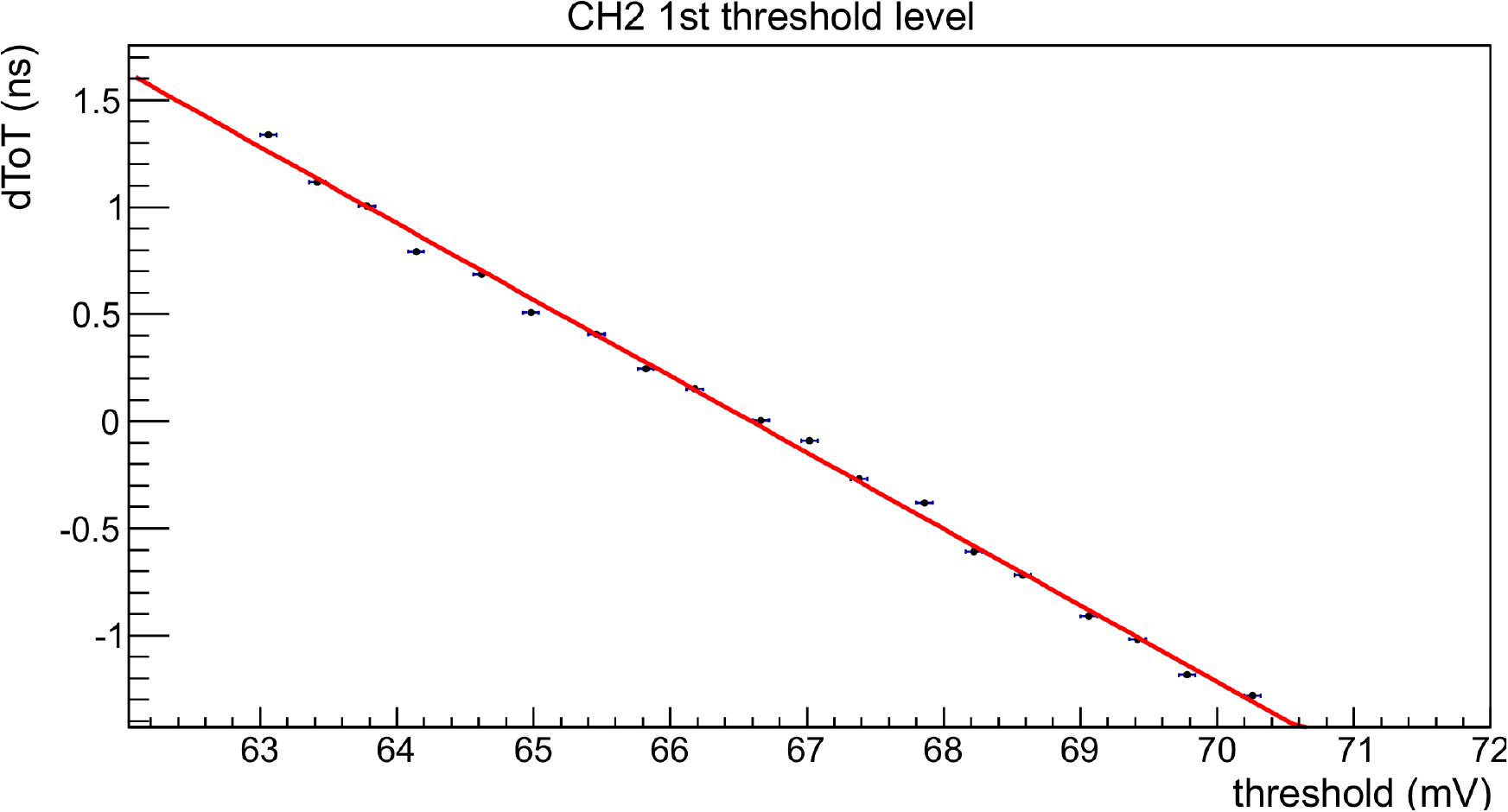}
	\caption{\label{fig:Fig11} Mean value of the distribution of dToT versus the employed threshold on the oscilloscope for the 2nd threshold of channel 1. The threshold value of the line intersection with the x axis, is the actual threshold value set on the card.}
\end{figure}

Subsequently the mean value of the dToT distributions is plotted against the employed threshold for the ToT calculation from the oscilloscope data and is depicted in Fig. ~\ref{fig:Fig10} for the 1st threshold of channel 1 of the card. The intersection of this line with the x axis corresponds to the actual threshold value set on the card. Figure ~\ref{fig:Fig11} displays the respective dToT plot versus the oscilloscope threshold for the 1st threshold of channel 2 of the card.

\begin{figure}[!t]
	\centering
	\includegraphics[width=4in]{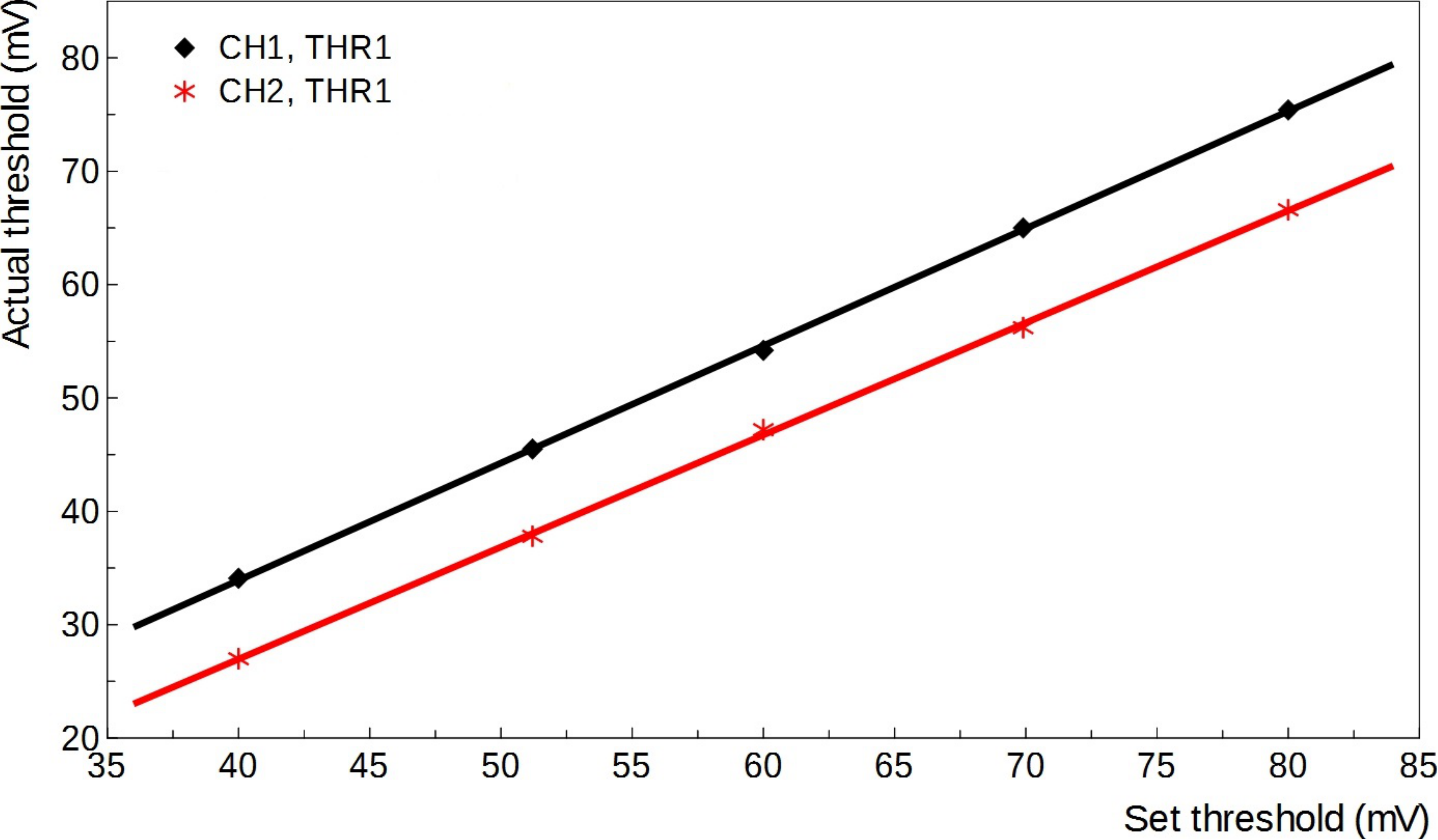}
	\caption{\label{fig:Fig12} Evaluated relation between set thresholds on the card and actual thresholds employed for the data acquisition, for the 1st threshold of channel 1 and channel 2. The line corresponds to a linear fit of the data and is used to determine the set threshold for a required actual threshold to be applied to the card at operation.}
\end{figure}

The calibration process is repeated for a number of set thresholds, evaluating each time the actual  threshold applied to the respective channel. The results for the 1st threshold of the two channels are depicted in Fig. ~\ref{fig:Fig12}. A linear fit of the data provides the relation between the set and actual value of each threshold, allowing thus the setting of the operating thresholds with high accuracy.

\begin{figure}[!t]
	\centering
	\includegraphics[width=4in]{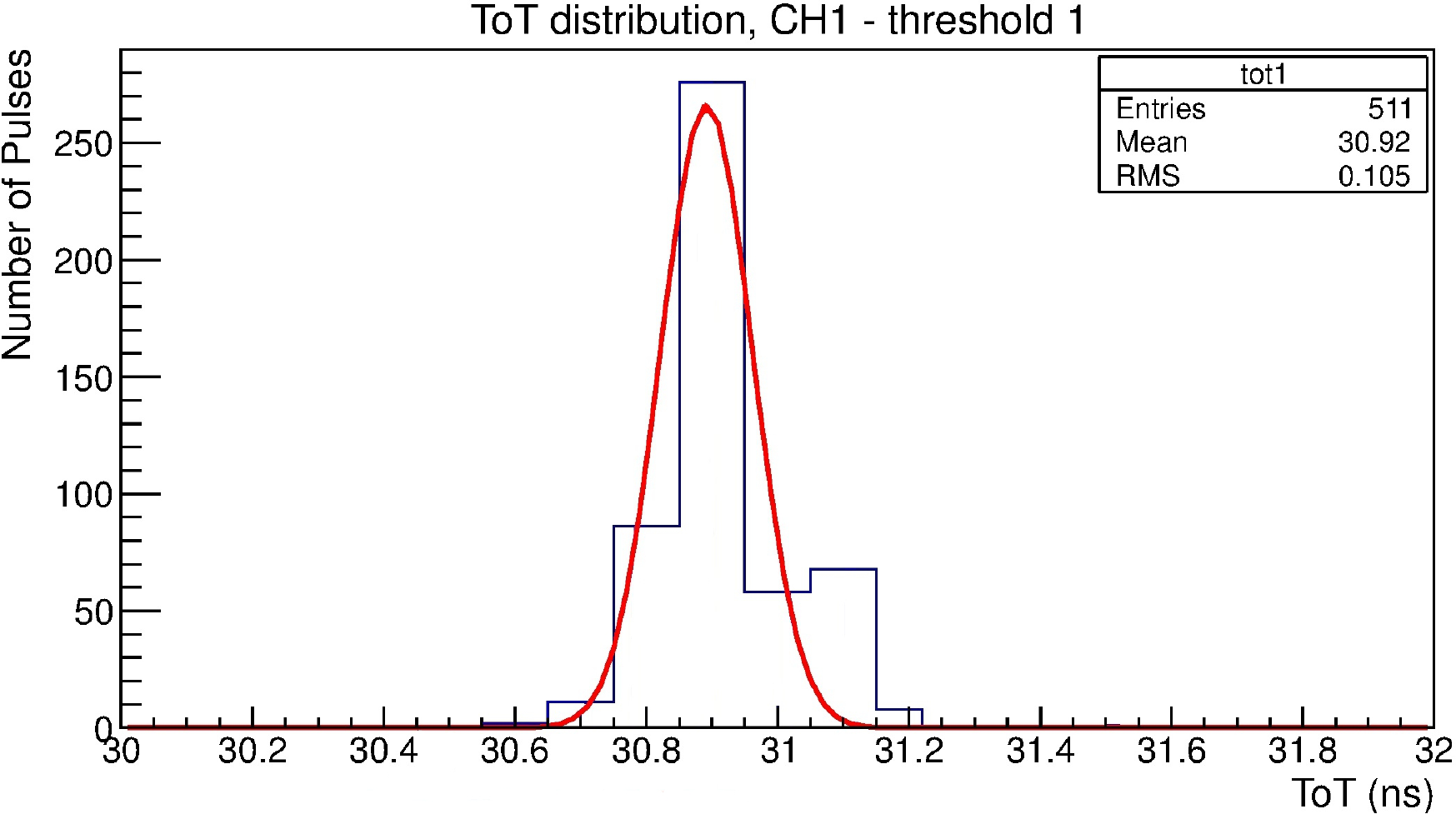}
	\caption{\label{fig:Fig13} Distribution of the ToT values for a certain 1st threshold value of the channel 1 of the card. The resolution of measuring ToT is 0.1 ns.}
\end{figure}

Finally, in Fig. ~\ref{fig:Fig13} the distribution of the ToT indicates that its resolution is around 0.1 ns, as expected, and so is the resolution in determining the leading edge of the pulse, i.e. the time of the first crossing of the pulse with the respective threshold level.

\section{Conclusion}
A data acquisition system has been presented that enables the measurement of PMT output pulses based on the MToT technique. The developed system consists of a prototype DAQ card, an FPGA development board and a GPS module. The measurement of the time is performed using a TDC which provides a less expensive and power consuming solution compared to a fast ADC, since high signal rate and short signals are detected. The design is characterized by a good calibration behavior and a resolution of 100 ps. The system performance has been evaluated and demonstrated by comparing its measurements to a commercial high-end, high-speed oscilloscope.

\section*{Acknowledgments}
This research has been co-financed by the European Union (European Social Fund - ESF) and Greek national funds through the Operational Program "Education and Lifelong Learning" of the National Strategic Reference Framework (NSRF) - Research Funding Program: "THALIS - HOU - Development and Applications of Novel Instrumentation and Experimental Methods in Astroparticle Physics".

\end{document}